\documentclass[runningheads,a4paper]{llncs}

\usepackage{amssymb}
\usepackage{graphicx}
\usepackage{color}
\usepackage{amsmath}
\usepackage{amssymb}
\usepackage{xspace}
\usepackage{enumerate}

\usepackage{url}
\urldef{\mailsa}\path|{alfred.hofmann, ursula.barth, ingrid.haas, frank.holzwarth,|
\urldef{\mailsb}\path|anna.kramer, leonie.kunz, christine.reiss, nicole.sator,|
\urldef{\mailsc}\path|erika.siebert-cole, peter.strasser, lncs}@springer.com|

\newcommand{\slength}[1]{|#1|}
\newcommand{\length}{\ensuremath{\mathit{len}}}
\newcounter{instr}
\newcommand{\ninstr}{\refstepcounter{instr}\theinstr.}

\newcommand{\ep}{end\text{-}pos}
\newcommand{\suf}{s\ell}  
\newcommand{\isuf}{s\ell^*}  
\newcommand{\aut}{\mathcal{A}}

\newcommand{\R}{\mathcal{R}\hspace{-1.2pt}_{_{x}}}
\newcommand{\notR}{\not\hspace{-3pt}\mathcal{R}\hspace{-2pt}_{_{x}}}
\newcommand{\bigO}{\mathcal{O}}
\newcommand{\Suff}{\mathit{Suff}\xspace}
\newcommand{\Fact}{\mathit{Fact}\xspace}
\newcommand{\myroot}{\mathit{root}\xspace}

\newcommand{\defAs}
{=}

\newcommand{\trdsuff}
{\stackrel{\textit{\tiny utd}}{\sqsupseteq}}

\newcommand{\val}{\mathit{val}\xspace}
\newcommand{\Prob}{\mathit{Pr}\xspace}

\newlength{\gnat}

\newlength{\gnatb}

\newcommand{\X}{X}
\newcommand{\Y}{Y}
\newcommand{\Z}{Z}

\newcommand{\expect}{E}

\newcommand{\md}{\mathit{\mu}\xspace}

\begin{document}
\mainmatter  

\title{Sequence Searching Allowing for Non-Overlapping Adjacent Unbalanced Translocations}

\titlerunning{Seaerching Allowing for Non-Overlapping Adjacent Unbalanced Translocations}
%
\author{Domenico Cantone\inst{1} 
\and
Simone Faro\inst{1} 
\and
Arianna Pavone\inst{2}
}
\authorrunning{D.Cantone, S.Faro and A.Pavone}
%
\institute{Dipartimento di Matematica e Informatica, Universit\`a di Catania,\\Viale Andrea Doria 6, I-95125 Catania, Italy \and
Dipartimento di Scienze Cognitive, Universit\`a di Messina,\\Via Concezione 6, 98122 Messina, Italy\\
\email{cantone@dmi.unict.it}\\
\email{faro@dmi.unict.it}\\
\email{apavone@unime.it}
}
\toctitle{Lecture Notes in Computer Science}
\tocauthor{Authors' Instructions}
\maketitle

\begin{abstract}
\emph{Unbalanced translocations} are among the most frequent chromosomal alterations, accounted for 30\% of all losses of heterozygosity, a major genetic event causing inactivation of tumor suppressor genes. Despite of their central role in genomic sequence analysis, little attention has been devoted to the problem of matching sequences allowing for this kind of chromosomal alteration.

In this paper we investigate the \emph{approximate string matching} problem when the edit operations are non-overlapping unbalanced translocations of adjacent factors.  In particular, we first present a $\bigO(nm^3)$-time and $\bigO(m^2)$-space algorithm based on the dynamic-programming approach.
Then we improve our first result by designing a second solution which makes use of the Directed Acyclic Word Graph of the pattern. In particular, we show that under the assumptions of equiprobability and independence of characters, our algorithm has a $\bigO(n\log^2_{\sigma} m)$ average time complexity, for an alphabet of size $\sigma$, still maintaining the $\bigO(nm^3)$-time and the $\bigO(m^2)$-space complexity in the worst case.
To the best of our knowledge this is the first solution in literature for the approximate string matching problem allowing for unbalanced translocations of factors.
\end{abstract}

\section{Introduction}\label{sec:Introduction}

Retrieving information and teasing out the meaning of biological sequences are central problems in modern biology.  Generally, basic biological information is stored in strings of nucleic acids (\textsc{dna}, \textsc{rna}) or amino acids (proteins).  


In recent years, much work has been devoted to the development of efficient methods for aligning strings and, despite sequence alignment seems to be a well-understood problem (especially
in the edit-distance model), the same cannot be said for the approximate string matching problem on biological sequences.

\emph{String alignment} and \emph{approximate string matching} are two fundamental problems in text processing. Given two input sequences $x$, of length $m$, and $y$, of length $n$, the \emph{string alignment} problem consists  in finding a set of edit operations able to transform $x$ in $y$, while the \emph{approximate string matching} problem consists in finding all approximate matches of $x$ in $y$.  The closeness of a match is measured in terms of the sum of the costs of the elementary edit operations necessary to convert the string into an exact match.

Most biological string matching methods are based on the \emph{Levenshtein distance}~\cite{Lev66}, commonly referred to just as \emph{edit distance}, or on the \emph{Damerau distance}~\cite{Dam64}. The edit operations in the case of the Levenshtein distance are \emph{insertions}, \emph{deletions}, and \emph{substitutions} of characters, whereas, in the case of the Damerau distance, \emph{swaps} of characters, i.e., transpositions of two adjacent characters, are also allowed (for an in-depth survey on approximate string matching, see~\cite{Nav01}).  Both distances assume that changes between strings occur locally, i.e., only a small portion of the string is involved in the mutation event.  However, evidence shows that in some cases large scale changes are possible~\cite{CH03,VAL06,CHK15} and that such mutations are crucial in \textsc{dna} since they often cause genetic diseases \cite{Lupski98,Oliver02}. 
For example, large pieces of \textsc{dna} can be moved from one location to another (\emph{translocations}) \cite{CHK15,WHR15,War91,OK08}, or replaced by their reversed complements (\emph{inversions}) \cite{CCF13}. 

Translocations can be \emph{balanced} (when equal length pieces are swapped) or \emph{unbalanced} (when pieces with different lengths are moved).
Interestingly, unbalanced translocations are a relatively common type of mutation and a major contributor to neurodevelopmental disorders \cite{WHR15}.
In addition, cytogenetic studies have also indicated that unbalanced translocations can be found in human genome with a de novo frequency of 1 in 2000 \cite{War91} and that it is a frequent chromosome alteration in a variety of human cancers \cite{OK08}. Hence the need for practical and efficient methods for detecting and locating such kind of large scale mutations in biological sequences.



\subsection{Related Results}
In the last three decades much work has been made for the alignment and matching problem allowing for chromosomal alteration, especially for non-overlapping inversions. Table \ref{tab:related} shows the list of all solutions proposed over the years, together with their worst-case, average-case and space complexities.

Concerning the alignment problem with inversions, a first solution based on dynamic programming, was proposed by Sch\"oniger and Waterman \cite{SW92}, which runs in $\bigO(n^2 m^2)$-time and $\bigO(n^2m^2)$-space on input sequences of length $n$ and $m$. Several other papers have been devoted to the alignment problem with inversions. The best solution is due to Vellozo \emph{et al.} \cite{VAL06}, who proposed a $\bigO(nm^2)$-time and $\bigO(nm)$-space algorithm, within the more general framework of an edit graph.

Regarding the alignment problem with translocations, Cho \emph{et al.} \cite{CHK15} presented a first solution for the case of inversions and translocations of equal length factors (i.e., balanced translocations), working in $\bigO(n^3)$-time and $\bigO(m^2)$-space. However their solution generalizes the problem to the case where edit operations can occur on both strings and assume that the input sequences have the same length, namely $|x|=|y|=n$.



Regarding the approximate string matching problem, a first solution was presented by Cantone \emph{et al.} \cite{CCF11}, where the authors presented an algorithm running in $\bigO(nm)$ worst-case time and $\bigO(m^{2})$-space for the approximate string matching problem allowing for non-overlapping inversions. Additionally, they also provided a variant \cite{CCF13} of the algorithm which has the same complexity in the worst case, but achieves $\bigO(n)$-time complexity on average.
Cantone \emph{et al.} also proposed in~\cite{CFG10} an efficient solution running in $\bigO(nm^2)$-time and $\bigO(m^2)$-space for a slightly more general problem, allowing for balanced translocations of adjacent factors besides non-overlapping inversions. The authors improved their previous result in~\cite{CFG14} obtaining an algorithm having $\bigO(n)$-time complexity on average.
We mention also the result by Grabowski \emph{et al.}~\cite{GFG11}, which solves the same string matching problem in $\bigO(nm^2)$-time and $\bigO(m)$-space, reaching in practical cases $\bigO(n)$-time complexity.


\begin{table}[t!]
\begin{footnotesize}
\begin{center}
\begin{tabular}{llllcr}
&   \textsc{Authors} &\textsc{Year} ~~~~& \textsc{W.C. Time} ~~&~~ \textsc{AVG Time}  &~~  \textsc{Space}\\
\hline
\multicolumn{6}{l}{\textbf{Alignment with inversions}}\\[0.2cm]
&   Schoniger and Waterman \cite{SW92} & (1992) & $\bigO(n^2m^2)$ & - & $\bigO(m^2)$\\
&   Gao \emph{et al.} \cite{Gao03} & (2003) & $\bigO(n^2m^2)$ & - & $\bigO(nm)$\\
&   Chen \emph{et al.} \cite{CGLNWW04}  & (2004) & $\bigO(n^2m^2)$ & - & $\bigO(nm)$\\
&   Alves \emph{et al.} \cite{ADV05}  & (2005) & $\bigO(n^3\log n)$ & - & $\bigO(n^2)$\\
&   Vellozo \emph{et al.} \cite{VAL06} & (2006) & $\bigO(nm^2)$ & - & $\bigO(nm)$\\[0.2cm]
\hline
\multicolumn{6}{l}{\textbf{Alignment with inversions and balanced translocations on both strings}}\\[0.2cm]
&   Cho \emph{et al.} \cite{CHK15} & (2015) & $\bigO(m^3)$ & - & $\bigO(m^2)$\\[0.2cm]
\hline
\multicolumn{6}{l}{\textbf{Pattern matching with inversions}}\\[0.2cm]				
&   Cantone \emph{et al.} \cite{CCF11} & (2011) & $\bigO(nm)$ & - & $\bigO(m^2)$\\
&   Cantone \emph{et al.} \cite{CCF13} & (2013) & $\bigO(nm)$ & $\bigO(n)$ & $\bigO(m^2)$\\[0.2cm]
\hline
\multicolumn{6}{l}{\textbf{Pattern matching with inversions and balanced translocations}}\\[0.2cm]
&   Cantone \emph{et al.} \cite{CFG10} & (2010) & $\bigO(nm^2)$ & $\bigO(n\log m)$ & $\bigO(m^2)$\\
&   Grabowski \emph{et al.} \cite{GFG11} & (2011) & $\bigO(nm^2)$ & $\bigO(n)$ & $\bigO(m)$\\
&   Cantone \emph{et al.} \cite{CFG14} & (2014) & $\bigO(nm^2)$ & $\bigO(n)$ & $\bigO(m)$\\[0.2cm]
\hline
\multicolumn{6}{l}{\textbf{Pattern matching with unbalanced translocations}}\\[0.2cm]
&   This paper & (2018) & $\bigO(nm^3)$ & $\bigO(n\log^2 m)$ & $\bigO(m^2)$\\[0.2cm]
\hline
\end{tabular}
\end{center}
\end{footnotesize}
\caption{\label{tab:related}Results related to alignment and matching of strings allowing for inversions and translocations of factors. All the edit operations allowed by all the listed solutions are intended to involve only non-overlapping factors of the pattern.}
\end{table}

\subsection{Our Results}
While in the previous results mentioned above it is intended that a translocation may take place only between balanced factors of the pattern,
in this paper we investigate the approximate string matching problem under a string distance whose edit operations are non-overlapping unbalanced translocations of adjacent factors. To the best of our knowledge, this slightly more general problem has never been addressed in the context of approximate pattern matching on biological sequences. Firstly, we propose a solution to the problem, based on the general dynamic programming approach, which needs $\bigO(nm^3)$-time and $\bigO(m^2)$-space.
Subsequently, we propose a second solution to the problem that makes use of the Directed Acyclic Word Graph of the pattern and achieves a $\bigO(n\log_{\sigma}^2 m)$-time complexity on average, for an alphabet of size $\sigma\geq 4$, still maintaining the same complexity, in the worst case, as for in the first solution.

The rest of the paper is organized as follows.  In Section~\ref{sec:Notions} we introduce some preliminary notions and definitions.  Subsequently, in Section~\ref{sec:newalgo1} we present
our first solution running in $\bigO(nm^3)$-time based on the dynamic programming approach.  In Section~\ref{sec:newalgo2} we present our second algorithm and analyze it both in the worst and in the average case.
Finally draw our conclusions in Section~\ref{sec:conclusions}.

\section{Basic notions and definitions}\label{sec:Notions}
A string $x$ of length $m \geq 0$, over an alphabet $\Sigma$, is represented as a finite array $x[1\,..\,m]$ of elements of $\Sigma$. We write $\slength{x} = m$ to indicate its length.  In particular, when $m=0$ we have the empty string $\varepsilon$.  We denote by $x[i]$ the $i$-th character of $x$, for $1\leq i\leq m$.  Likewise, the
substring of $x$ factor, contained between the $i$-th and the $j$-th characters of $x$ is indicated with $x[i\,..\,j]$, for $1\leq i \leq j \leq m$.  The set of factors of $x$ is
denoted by $\Fact(x)$ and its size is $\bigO(m^2)$.  

A string $w\in \Sigma^*$ is a suffix of $x$ (in symbols, $w \sqsupseteq x$) if $w = x[i\,..\,m]$, for some $1\leq i \leq m$. We denote by $\Suff(x)$ the set of the suffixes of $x$.  Similarly, we say that $w$ is a prefix of $x$ if $w = x[1\,..\,i]$, for some $1\leq i \leq m$.  Additionally, we use the symbol $x_i$ to denote the prefix of $x$ of length $i$ (i.e.,$x_i = x[1\,..\,i]$), for $1\leq i\leq m$, and make the convention that $x_{0}$ denotes the empty string $\varepsilon$.  In addition, we write $xw$ for the concatenation of the strings $x$ and $w$.

For $w\in \Fact(p)$, we denote with $\ep(w)$ the set of all positions in $x$ at which an occurrence of $w$ ends; formally, we put $\ep(w) := \{i\ |\ w \sqsupseteq x_{i}\}$.  For any given pattern $x$, we define an equivalence relation $\R$ 
by putting, for all $w,z\in \Sigma^*$, 
$$w \mathrel{\R} z \iff \ep(w)=\ep(z).$$ 
We also denote with $\R(w)$ the equivalence class 
of the string $w$. For each equivalence class $q$ of $\R$, we put $\length(q) = |\val(q)|$, where $\val(q)$ is the longest string $w$ in the
equivalence class $q$. 

\begin{example}\label{ex1}
Let $x=agcagccag$ be a string over $\Sigma=\{a,g,c,t\}$ of length $m=9$.  Then we have $\ep(ag)=\{2,5,9\}$, since the substring $ag$ occurs three times in $x$, ending at positions $2$, $5$ and $9$, respectively, in that order. Similarly we have $\ep(gcc)=\{7\}$.
Observe that $\R(ag)=\{ag,g\}$.  Similarly we have
$\R(gc)=\{agc,gc\}$.  Thus, we have $\val(\R(ag))=ag$, $\length(\R(ag))=2$, $\val(\R(gc))=agc$ and $\length(\R(gc))=3$.
\end{example}

The Directed Acyclic Word Graph~\cite{CR94} of a pattern $x$ (\textsc{Dawg}, for short) is the deterministic automaton $\mathcal{A}(x)=(Q,\Sigma,\delta,\myroot,F)$ whose language is $\Fact(x)$, where $Q=\{\R(w) : w\in \Fact(x)\}$ is the set of states, $\Sigma$ is the alphabet of the characters in $x$, $\myroot=\R(\varepsilon)$ is the initial state, $F=Q$ is the set of final states, and $\delta:Q\times \Sigma\rightarrow Q$ is the transition function defined by $\delta(\R(y),c) \defAs \R(yc)$, for all $c\in \Sigma$ and
$yc\in \Fact(x)$.

We define a failure function, $\suf:\Fact(x) \setminus \{\varepsilon\} \rightarrow \Fact(x)$, called \emph{suffix link}, by putting, for any $w \in \Fact(x) \setminus \{\varepsilon\}$,
$$\suf(w) = \text{``\,longest } y \in \Suff(w) \text{ such that } y \notR w\text{"}.$$
The function $\suf$ enjoys the following property $$w \mathrel{\R} y \Longrightarrow \suf(w)=\suf(y).$$  We extend the functions $\suf$ and $\ep$ to $Q$ by putting $\suf(q) := \R(\suf(\val(q)))$ and $\ep(q) = \ep(\val(q))$, for each $q \in Q$. Figure \ref{fig:dawg} shows the \textsc{Dawg} of the pattern $x=aggga$, where the edges of the automaton are depicted in black while the suffix links are depicted in red.

\smallskip

A \emph{distance} $d:\Sigma^{*}\times \Sigma^{*}\rightarrow \mathbb{R}$ is a function which associates to any pair of strings $x$ and $y$ the minimal cost of any finite sequence of edit operations which transforms $x$ into $y$, if such a sequence exists, $\infty$ otherwise.  

In this paper we consider the \emph{unbalanced translocation distance}, $utd(x,y)$, whose unique edit operation is the \emph{translocation} of two adjacent factors of the string, with possibly different lengths.
Specifically, in an \emph{unbalanced translocation} a factor of the form $zw$ is transformed into $wz$, provided that both $\slength{z},\slength{w} > 0$ (it is not necessary that $\slength{z}=\slength{w}$). We assign a unit cost to each translocation.

\begin{example}\label{ex2}
Let $x=g\overline{t}\underline{ga}c\overline{cgt}\underline{ccag}$ and $y=g\underline{ga}\overline{t}c\underline{ccag}\overline{cgt}$ be given two strings of length $12$.  Then $utd(x,y)=2$ since $x$ can be transformed into $y$ by translocating the substrings $x[3..4]=ga$ and $x[2..2]=t$, and translocating the substrings $x[6..8]=cgt$ and $x[9..12]=ccag$.
\end{example}

When $utd(x,y) <\infty$, we say that $x$ and $y$ have $utd$-match. If $x$ has $utd$-match with a suffix of $y$, we write $x \trdsuff y$.


\section{A Dynamic Programming Solution}
\label{sec:newalgo1}
In this section we present a general dynamic programming algorithm for the pattern matching problem with adjacent unbalanced translocations. 

Let $x$ be a pattern of length $m$ and $y$ a text of length $n$, both over the same alphabet $\Sigma$ of size $\sigma$.
Our algorithm is designed to iteratively compute, for $j = m,m+1,\ldots,n$, all the prefixes of $x$ which have a $utd$-match with the suffix  $y_{j}$, by exploiting information gathered at previous iterations.
For this purpose, it is convenient to introduce the set $\texttt{P}_j$, for $1 \leq j \leq n$, defined by  
$$\texttt{P}_j := \{1\leq i\leq m\ |\ x_i \trdsuff y_j\}.$$
Thus, the pattern $x$ has an $utd$-match ending at position $j$ of the text $y$ if and only if $m\in \texttt{P}_j$.

Since the allowed edit operations involve substrings of the pattern $x$, it is useful to consider also the set $\texttt{F}^k_j$ of all the positions in $x$ at which an occurrence of the suffix of $y_{j}$ of length $k$ ends.  More precisely, for $1\le k \leq m$ and $k-1\leq j \leq n$, we put 
$$ \texttt{F}^k_j := \{k-1\leq i\leq m\ |\ y[j-k+1\,..\,j] \sqsupseteq x_{i}\}\,.$$
Observe that 
\begin{equation}\label{obs1}
\texttt{F}^k_j \subseteq \texttt{F}^h_j, \textrm{ for } 1\leq h \leq k \leq m.
\end{equation}

\begin{example}\label{ex3}
Let $x=cattcatgatcat$ $y=atcatgacttactgactta$ be a pattern and respectively
a text.  Then $\texttt{F}^3_5$ is the set of all positions in $x$ at which an occurrence of the suffix of $y_{5}$ of length $3$ ends,
namely, $cat$.  Thus $\texttt{F}^3_5 = \{3,7,13\}$.  Similarly, we have that $\texttt{F}^2_5 = \{3,7,10,13\}$.  Observe that $\texttt{F}^3_5 \subseteq \texttt{F}^2_5$.
\end{example}

The sets $\texttt{P}_j$ and $\texttt{F}^k_j$ can then be computed by way of the recursive relations contained in the following elementary lemmas.

\begin{lemma}\label{lem1}
Let $y$ and $x$ be a text of length $n$ and a pattern of length $m$, respectively.  Then $i \in \texttt{P}_{j}$, for $1\leq i \leq m$ and $i\leq j\leq n$,
if and only if one of the following two facts holds:
\begin{enumerate}[(a)]
    \item \label{first}
    $x[i]= y[j]$ and
    $(i - 1) \in \texttt{P}_{j-1} \cup \{0\}$;

    \item  \label{second}
    $(i-k)\in \texttt{F}^h_j$,  $i\in  \texttt{F}^k_{j-h}$, and $(i - k-h) \in \texttt{P}_{j-k-h} \cup \{0\}$,
    for some $1 \leq k,h < i$ such that $h+k\leq i$;\qed
\end{enumerate}
\end{lemma}
Notice that condition (\ref{second}) in Lemma~\ref{lem1} refers to a translocation of adjacent factors of length $k$ and $h$, respectively.

\begin{figure}[t]
\begin{center}
\setlength{\unitlength}{0.01\textwidth}
\setlength{\fboxrule}{1mm}
\begin{picture}(84,27)

 \put(0,18){\makebox(30,3)[l]{$y$}}
 \put(3,18){  $\underbrace{\framebox(21,4)[c]{...}
     		                         \framebox(14,4)[c]{$s$}
		                         \hspace{-1pt}\overbrace{\framebox(16,4)[c]{$w$}}^{k}
		                         \hspace{-1pt}\overbrace{\framebox(10,4)[c]{$z$}}^{h}
		                        }_{j}
		                        \hspace{-1pt}\framebox(21,4)[c]{...}$ 
		}

 \put(22,4){\makebox(30,3)[l]{$x$}}
 \put(24,4){ $\underbrace{
 			\framebox(14,4)[c]{$u$}
                         \hspace{-1pt}\overbrace{\framebox(10,4)[c]{$z$}}^{h}
                         \hspace{-1pt}\overbrace{\framebox(16,4)[c]{$w$}}^{k}
                       }_{i}$
                       \hspace{-3pt}\framebox(8,4)[c]{}
                 }

\end{picture}
\end{center}
\caption{\label{fg1}Case (b) of Lemma \ref{lem1}. The prefix $u$ of the pattern, of length $i-h-k$, has a $utd$-match ending at position $j-h-k$ of the text, i.e. $(i-h-k)\in \texttt{P}_{j-k-h} \cup \{0\}$. In addition the substring of the pattern $z = x[i-h-k+1..i-k]$, of length $h$ has an exact match with the substring of the text $y[j-h+1..j]$, i.e. $(i-k)\in \texttt{F}^h_j$. Finally the substring of the pattern $w = x[i-k+1..i]$, of length $k$ has an exact match with the substring of the text $y[j-h-k+1..j-h]$, i.e. $i \in \texttt{F}^k_{j-h}$.}
\end{figure}
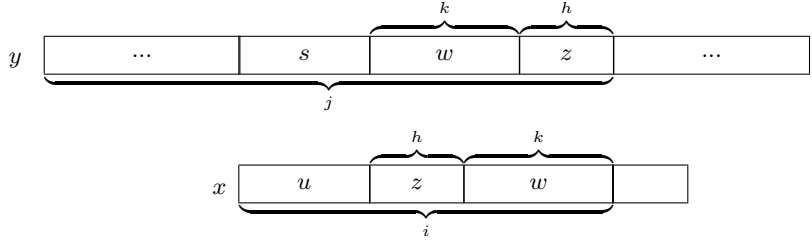

Likewise, the sets $\texttt{F}^k_j$ can be computed according to the following lemma.

\settowidth{\gnat}{$(k=1 \textrm{ or }i\in \mathcal{I}^{k-1}_{j-1}) \textrm{ and } p[i-k+1]=y[j]\,.$}
\settowidth{\gnatb}{$\sqcup$}
\setlength{\gnat}{-\gnat}
\addtolength{\gnat}{\textwidth}
\addtolength{\gnat}{-\gnatb}

\begin{lemma}\label{lem2}
Let $y$ and $x$ be a text of length $n$ and a pattern of length $m$, respectively.  Then $i \in \texttt{F}^k_{j}$ if and only if one of the following condition holds:
\begin{enumerate}[(a)]
\item $x[i]=y[j]$ and $k=1$;
\item $x[i]=y[j]$, $1<k<i$ and $(i-1)\in \texttt{F}^{k-1}_{j-1}$,
\end{enumerate}
for $1\le k< i < m$ and $k-1\leq j\leq n$. \qed
\end{lemma}

\begin{figure}[!t]
\begin{center}
\begin{tabular}{rl}
\multicolumn{2}{l}{~~\textsc{Algorithm1}}\\
~\textsf{1.} & \textsf{$\texttt{P}[0,0] \leftarrow$ true}\\ 
~\textsf{2.} & \textsf{for $j \leftarrow 1$ to $n$ do}\\ 
~\textsf{3.} & \quad \textsf{$\texttt{P}[0,j] \leftarrow$ true}\\ 
~\textsf{4.} & \quad \textsf{for $i \leftarrow 1$ to $m$ do}\\ 
~\textsf{5.} & \quad \quad \textsf{if ($x[i]=y[j]$) then}\\ 
~\textsf{6.} & \quad \quad \quad \textsf{if ($\texttt{P}[i-1,j-1]$) then}\\ 
~\textsf{7.} & \quad \quad \quad \quad \textsf{$\texttt{P}[i,j]\leftarrow$ true}\\ 
~\textsf{8.} & \quad \quad \quad \textsf{$\texttt{F}[i,j] \leftarrow \texttt{F}[i-1,j-1]+1$}\\ 
~\textsf{9.} & \quad \quad \textsf{for $k \leftarrow 1$ to $i-1$ do}\\ 
~\textsf{10.} & \quad \quad \quad \textsf{for $h \leftarrow 1$ to $i-k$ do}\\ 
~\textsf{11.} & \quad \quad \quad \quad \textsf{if ($\texttt{F}[i-h,j]\geq k$ and $\texttt{F}[i,j-k]\geq h$) then}\\ 
~\textsf{12.} & \quad \quad \quad \quad \quad \textsf{if ($\texttt{P}[i-h-k,j-h-k]$) then}\\ 
~\textsf{13.} & \quad \quad \quad \quad \quad \quad \textsf{$\texttt{P}[i,j]\leftarrow$ true}\\ 
\end{tabular}
\caption{\label{fig:code1}A dynamic programming algorithm for the approximate string matching problem allowing for unbalanced translocations of adjacent factors. The algorithm is characterised  by a $\bigO(nm^3)$-time and a $\bigO(m^2)$-space complexity.}
\end{center}
\end{figure}

Based on the recurrence relations in Lemmas~\ref{lem1} and~\ref{lem2}, a general dynamic programming algorithm can be readily constructed, characterized by an overall $\bigO(n m^3)$-time and $\bigO(m^3)$-space complexity.  However, the overhead due to the computation and the maintenance of the sets $\texttt{F}^{k}_{j}$ turns out to be quite relevant.

Based on equation (\ref{obs1}), we can represent the sets $\texttt{F}^{k}_{j}$ by means of a single matrix $\texttt{F}$ of size $m^2$. Specifically, for $1\leq i\leq m$ and $1\leq j \leq n,$ we set $\texttt{F}[i,j]$ as the length of the longest suffix of $x_i$ which is also a suffix of $y_j$. More formally we have:
$$
	i \in \texttt{F}^{k}_{j} \iff \texttt{F}[i,j]\geq k.
$$
The pseudocode of the resulting dynamic programming algorithm, called \textsc{Algorithm1}, is shown in Figure \ref{fig:code1}. Due to the four for-loops at lines $2$, $4$, $9$, and $10$, respectively, it can straightforwardly be proved that \textsc{Algorithm1} has a $\bigO(n m^3)$-time and $\bigO(m^2)$-space complexity.
Indeed, the matrices $\texttt{F}$ and $\texttt{P}$ are filled by columns and therefore we need to store only the last $m$ columns at each iteration of the for-loop at line $2$.


\section{An Automaton-Based Algorithm}\label{sec:newalgo2}
In this section we improve the algorithm described in Section 3 by means of an efficient method for computing the sets $\texttt{F}^{k}_{j}$, for $1 \leq j \leq n$ and $1 \leq k < j$. Such method makes use of the \textsc{Dawg} of the pattern $x$ and the function $\ep$. 

Let $\mathcal{A}(x)=(Q, \Sigma, \delta, \myroot, F)$ be the \textsc{Dawg} of $x$.  For each position $j$ in $y$, let $w$ be the longest factor of
$x$ which is a suffix of $y_j$ too; also, let $q_j$ be the state of $\mathcal{A}(x)$ such that $\R(w)=q_j$, and let $l_j$ be the length of $w$.  We call the pair $(q_j,l_j)$ a
\emph{$y$-configuration} of $\mathcal{A}(x)$.  

The idea is to compute the $y$-configuration $(q_j, l_j)$ of $\mathcal{A}(x)$, for each position $j$ of the text, while scanning the text $y$.  The set $\texttt{F}^{k}_{j}$ computed at previous iterations do not need to be maintained explicitly; rather, it is enough to maintain only $y$-configurations.  These are then used to compute efficiently the set $\texttt{F}^{k}_{j}$ only when needed.

\begin{figure}
\begin{center}
\includegraphics[width=0.6\textwidth]{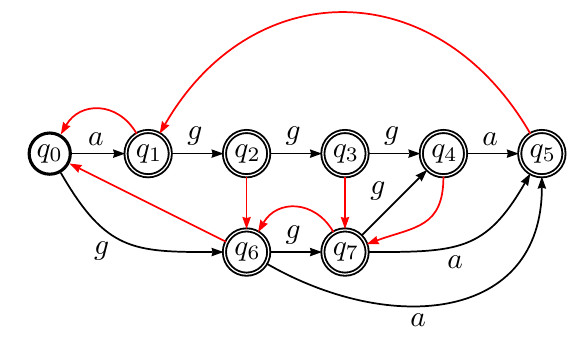}
\end{center}
\caption{\label{fig:dawg}The Directed Acyclic Word Graph (\textsc{Dawg}) of the pattern $x=aggga$, where the edges of the automaton are depicted in black while the suffix links are depicted in red.}
\end{figure}

\begin{example}\label{ex4}
Let $x=aggga$ be a pattern and $y=aggagcatgggactaga$ a text respectively. Let $\mathcal{A}(x)=(Q, \Sigma, \delta, \myroot, F)$ be the DAWG of $x$ as depicted in Figure \ref{fig:dawg}, where $\myroot=q_0$ is the initial state and $F=\{q_1,q_2,q_3,q_4,q_5,q_6,q_7\}$ is the set of final states.  Edges of the \textsc{Dawg} are depicted in black while suffix links are depicted in red.
Observe that, after scanning the suffix $y_6$ starting from state $q_0$ of $\mathcal{A}(x)$, we reach state $q_2$ of the automaton. Thus, the corresponding $y$-configuration is $(q_2,2)$. Similarly, after scanning the suffix $y_{11}$, we get the $y$-configuration $(q_4,3)$.
\end{example}

The longest factor of $x$ ending at position $j$ of $y$ is computed in the same way as in the Forward-Dawg-Matching algorithm for the exact pattern matching problem (the interested readers are referred to \cite{CR94} for further details).

Specifically, let $(q_{j-1},l_{j-1})$ be the $y$-configuration of $\mathcal{A}(x)$ at step $(j-1)$.
The new $y$-configuration $(q_j, l_j)$ is set to $(\delta(q,y[j]), length(q)+1)$, where $q$ is the first node in the suffix path $\langle q_{j-1}, \suf(q_{j-1}), \suf^{(2)}(q_{j-1}),\ldots\rangle$ of $q_{j-1}$, including $q_{j-1}$, having a transition on $y[j]$, if such a node exists; otherwise $(q_j, l_j)$ is set to $(\myroot,0)$.\footnote{We recall that $\suf^{(0)}(q) \defAs q$ and,  recursively, $\suf^{(h+1)}(q) \defAs \suf(\suf^{(h)}(q))$, for  $h\geq 0$. 
}

\smallskip

Before explaining how to compute the sets $\texttt{F}^{k}_{j}$, it is convenient to introduce the partial function $\phi:Q \times \mathbb{N} \rightarrow Q$, which, given a node
$q \in Q$ and a length $k \leq length(q)$, computes the state $\phi(q,k)$ whose corresponding set of factors contains the suffix of $\val(q)$ of length $k$.  
Roughly speaking, $\phi(q,k)$ is the first node $p$ in the suffix path of $q$ such that $length(\suf(p)) < k$.

In the preprocessing phase, the \textsc{Dawg} $\mathcal{A}(x)=(Q, \Sigma,\delta, \myroot, F)$, together with the associated $\ep$ function, is computed.  Since for a pattern $x$ of length $m$ we have $|Q| \leq 2m+1$ and $|\ep(q)| \leq m$, for each $q \in Q$, we need only $\bigO(m^2)$ extra space (see \cite{CR94}).

To compute the set $\texttt{F}^k_j$, for $1\le k\le l_j$, we take advantage of the relation
\begin{equation}
    \label{eq:fac}
    \texttt{F}^k_j=\ep(\phi(q_j,k)).
\end{equation}
Notice that, in particular, we have $\texttt{F}^{l_j}_j=\ep(q_j)$.

The time complexity of the computation of $\phi(q,k)$ can be bounded by the length of the suffix path of node $q$.  Specifically, since the sequence
$$\langle\length(\suf^{(0)}(q)),\length(\suf^{(1)}(q)), \ldots, 0\rangle$$
of the lengths of the nodes in the suffix path from $q$ is strictly decreasing, we can do at most $\length(q)$ iterations over the suffix link, obtaining a $\bigO(m)$-time complexity.

According to Lemma~\ref{lem1}, a translocation of two adjacent factors of length $k$ and $h$, respectively, at position $j$ of the text $y$ is possible only if two factors of $x$ of lengths at least $k$ and $h$, respectively, have been recognized at both positions $j$ and $j-h$, namely if $l_j\ge h$ and $l_{j-h}\ge k$ hold (see Figure \ref{fg1}).

Let $\langle h_1, h_2, \ldots, h_r \rangle$ be the increasing sequence of all the values $h$ such that $1\leq h \leq \min(l_{j},l_{j-h})$.  For each $1\leq i \leq r$, condition (\ref{second}) of Lemma~\ref{lem1} requires member queries on the sets $\texttt{F}^{h_i}_{j}$.

Observe that if we proceed for decreasing values of $h$, the sets $\texttt{F}^{h}_{j}$, for $1 \leq h \leq l_j$, can be computed in constant time.  Specifically, for $h=1, \ldots, l_j-1$ $\texttt{F}^h_j$ can be computed in constant time from $\texttt{F}^{h+1}_j$ with at most one iteration over the suffix link of the state $\phi(q_{j},h+1)$.

Subsequently, for each member query on the set $\texttt{F}^{h_i}_{j}$, condition (\ref{second}) of Lemma~\ref{lem1} requires also member queries on the sets $\texttt{F}^{k}_{j-{h_i}}$, for $1\leq k < h_i$. Let $\langle k_1, k_2, \ldots, k_s \rangle$ be the increasing sequence of all the values $k$ such that $1\leq k \leq \min(l_{j-h_i},l_{j-h_i-k})$.
Also in this case we can proceed for decreasing values of $k$, in order to compute the sets $\texttt{F}^{k}_{j-{h_i}}$ in constant time, for $1 \leq k \leq l_{j-h_i}$. 
\smallskip

The resulting algorithm for the approximate string matching problem allowing for unbalanced translocations of adjacent factors is shown in Figure \ref{fig:code2} and is named \textsc{Algorithm2}. In the next sections, we analyze its worst-case and average-case complexity.

\begin{figure}[t!]
\begin{tabular}{rl}
\multicolumn{2}{l}{~~\textsc{Algorithm2}}\\
~\textsf{\ninstr} & \textsf{$m \leftarrow \slength{x}$, \quad $n \leftarrow \slength{y}$}\\
~\textsf{\ninstr} & \textsf{$(q_0,l_0) \leftarrow\textsc{Dawg-Delta}(root_{\aut},0,y[0],\aut)$}\\
~\textsf{\ninstr} & \textsf{$\texttt{P}_{0} \leftarrow \{0\}$}\\
~\textsf{\ninstr} & \textsf{\label{l:main}  for $j \leftarrow 1$ to $n$ do}\\
~\textsf{\ninstr} & \quad \textsf{$(q_j,l_j) \leftarrow\textsc{Dawg-Delta}(q_{j-1},l_{j-1},y[j],\aut)$}\\
~\textsf{\ninstr} & \quad \textsf{$\texttt{P}_{j} \leftarrow \{0\}$}\\
~\textsf{\ninstr} & \textsf{\label{loop:2}  \quad for $i\in \texttt{P}_{j-1}$ do}\\
~\textsf{\ninstr} & \quad \quad \textsf{if $i< m$ and $x[i+1]=y[j]$ then} \\
~\textsf{\ninstr} & \quad \quad \quad \textsf{$\texttt{P}_{j} \leftarrow \texttt{P}_{j} \cup \{i+1\}$}\\
~\textsf{\ninstr} & \quad \textsf{$u\leftarrow q_j$}\\
~\textsf{\ninstr} & \quad \textsf{\label{loop:3}for $h\leftarrow l_j$ downto $1$ do}\\
~\textsf{\ninstr} & \quad \quad \textsf{if $h = \length(\suf_{\aut}(u))$ then $u\leftarrow \suf_{\aut}(u)$}\\
~\textsf{\ninstr} & \quad \quad \textsf{$p\leftarrow q_{j-h}$}\\
~\textsf{\ninstr} & \quad \quad \textsf{\label{loop:4}for $k\leftarrow l_{j-h}$ downto $1$ do}\\
~\textsf{\ninstr} & \quad \quad \quad \textsf{if $k = \length(\suf_{\aut}(p))$ then $p\leftarrow \suf_{\aut}(p)$}\\
~\textsf{\ninstr} & \quad \quad \textsf{\label{loop:5}  \quad for $i\in \texttt{P}_{j-h-k}$ do}\\
~\textsf{\ninstr} & \quad \quad \quad \quad \textsf{\label{l:end}if ($(i+h)\in \ep(u)$ and $(i+h+k)\in \ep(p))$ then}\\
~\textsf{\ninstr} & \quad \quad \quad \quad \quad \textsf{$\texttt{P}_j\leftarrow \texttt{P}_j\cup \{i+h+k\}$}\\
~\textsf{\ninstr} & \quad \textsf{if $(m \in \texttt{P}_{j})$ then Output($j$)}\\
&\\
\multicolumn{2}{l}{~~\textsc{Dawg-Delta}$(q, l, c,\mathcal{A})$}\\
\setcounter{instr}{0}
~\textsf{\ninstr} & \textsf{if $\delta_{\mathcal{B}}(q, c) = \textsc{nil}$ then}\\
~\textsf{\ninstr} & \quad \textsf{do}\\
~\textsf{\ninstr} & \quad \quad \textsf{$q\leftarrow \isuf_{\mathcal{A}}(q)$}\\
~\textsf{\ninstr} & \quad \textsf{while $q\neq \textsc{nil}$ and $\delta_{\mathcal{A}}(q, c) = \textsc{nil}$}\\
~\textsf{\ninstr} & \quad \textsf{if $q = \textsc{nil}$ then}\\
~\textsf{\ninstr} & \quad \quad \textsf{$l\leftarrow 0,q\leftarrow \myroot_{\mathcal{A}}$}\\
~\textsf{\ninstr} & \quad \textsf{else $l\leftarrow \length(q)+1$}\\
~\textsf{\ninstr} & \quad \textsf{\phantom{else} $q\leftarrow \delta_{\mathcal{A}}(q,c)$}\\
~\textsf{\ninstr} & \textsf{else $l\leftarrow l+1$}\\
~\textsf{\ninstr} & \textsf{\phantom{else} $q\leftarrow \delta_{\mathcal{A}}(q, c)$}\\
~\textsf{\ninstr} & \textsf{return $(q,l)$}\\
\end{tabular}
\caption{\label{fig:code2}The code of \textsc{Algorithm2} and the \textsc{Dawg} state update algorithm \textsc{Dawg-Delta}.
Notice that the function $\isuf$ in procedure \textsc{Dawg-Delta} denotes the improved suffix link~\cite{CR94}.}
\end{figure}


\subsection{Worst-case time and space analysis}

In this section we determine the worst-case time and space analysis of \textsc{Algorithm2} presented in the previous section.  In particular, we will refer to the implementation of the          \textsc{Algorithm2} reported in Figure~\ref{fig:code2}.

First of all, observe that the main for-loop at line \ref{l:main} is always executed $n$ times.  For each of its iterations, the cost of the execution \textsc{Dawg-Delta} (line $5$) for computing the new $y$-configuration requires at most $\bigO(m)$-time.  Since we have $|\texttt{P}_j|\leq m+1$, for all $1\leq j \leq n$, the for-loop al line 7 is also executed $\bigO(m)$-times. In addition, since we have $l_j\leq m$, for all $1\leq j \leq n$, the two nested for-loops at lines $11$ and $14$ are executed $m$ times. Observe also that the transitions of suffix links performed at lines $12$ and $15$ need only constant time. Thus, at each iteration of the main for-loop, the internal for-loop at lines 14 takes at most $\bigO(m)$-time, while the for-loop at line 11 takes at most $\bigO(m^2)$-time.  In addition the for-loop at line 16 takes at most $\bigO(m)$-time, since $|\texttt{P}_{j-h-k}|\leq m+1$ and the tests at line 17 can be performed in constant time.
Summing up, \textsc{Algorithm2} has a $\bigO(nm^3)$ worst-case time complexity.

In order to evaluate the space complexity of \textsc{Algorithm2}, we observe that in the worst case, during the $j$-th iteration of its main for-loop, the sets $\texttt{F}^k_{j-k}$ and $\texttt{P}_{j-k}$, for $1\leq k \leq m$, must be kept in memory to handle translocations.
However, as explained before, we do not keep the values of
$\texttt{F}^k_{j-k}$ explicitly but rather we maintain only their corresponding $y$-configurations of the automaton $\mathcal{A}(x)$.
Thus, we need $\bigO(m)$-space for the last $m$ configurations of the automaton and $\bigO(m^2)$-space to keep the last $m+1$ values of the sets $\texttt{P}_{j-k}$, since the maximum cardinality of each such set is $m+1$.  Observe also that, although the size of the \textsc{Dawg} is linear in $m$, the \emph{end\textendash pos}$(\cdot)$ function can require $\bigO(m^2)$-space.  Therefore, the total space complexity of the \textsc{Algorithm2} is $\bigO(m^2)$.

\subsection{Average-case time analysis}
\label{av_case}

In this section we evaluate the average time complexity of our new automaton-based algorithm, assuming the uniform distribution and independence of characters in an alphabet $\Sigma$ with $\sigma\geq 4$ characters. 

In our analysis we do not include the time required for the computation of the DAWG and the \emph{end\textendash pos}$(\cdot)$ function, since they require $\bigO(m)$ and $\bigO(m^2)$ worst-case time, respectively, which turn out to be negligible if we assume that $m$ much smaller than $n$. Hence we evaluate only the searching phase of the algorithm. 

\newcommand{\Xj}[1]{\X(#1)}
\newcommand{\Yj}[1]{\Y(#1)}
\newcommand{\Zj}[1]{\Z(#1)}

Given an alphabet
$\Sigma$ of size $\sigma \geq 4$, for $j=1,\ldots,n$, we consider
the following nonnegative random variables over the sample space of
the pairs of strings $x,y \in \Sigma^{*}$ of length $m$ and $n$,
respectively:
\begin{itemize}
\item $\Xj{j} \defAs $ the length $l_j\le m$ of the longest factor of $x$ which is a 	suffix of $y_{j}$;
\item $\Zj{j} \defAs |\texttt{P}_j|$, where we recall that $\texttt{P}_j = \{1\leq i\leq m\ |\ x_i \trdsuff y_j\}$.
\end{itemize}
Then the run-time of a call to \textsc{Algorithm2} with parameters $(x,y)$ is proportional to 
\begin{equation} \label{sums}
\sum_{j=1}^{n}\left( \Zj{j-1} + \Xj{j} + \sum_{h=1}^{\Xj{j}} \left( \Xj{j-h} +  \sum_{k=1}^{\Xj{j-h}} \Zj{j-h-k} \right)\!\right),
\end{equation}
where the external summation refers to the main for-loop (at line 4), the second summation within it takes care of the internal for-loop at line 11, and the third summation refers to the inner for-loop at line 14.

Hence the average-case complexity of \textsc{Algorithm2} is the expectation of (\ref{sums}), which, by linearity, is equal to

\begin{footnotesize}
\begin{equation}\label{sumsb}
\sum_{j=1}^{n}\!\!\left(\! \expect(\Zj{j-1}) + \expect( \Xj{j}) + \expect\left(\! \sum_{h=1}^{\Xj{j}} \Xj{j-h} \!\right)+ \expect\left(\! \sum_{h=1}^{\Xj{j}}\sum_{k=1}^{\Xj{j-h}} \Zj{j-h-k}\!\right)  \!\right) \!\!.
\end{equation}
\end{footnotesize}

where $\expect(\cdot)$ be the expectation function. Since $\expect(\Xj{j}) \leq \expect(\Xj{n-1})$ and $\expect(\Zj{j}) \leq \expect(\Zj{n-1})$, for $1 \leq j \leq n$,\footnote{In fact, for $j=m+1,\ldots,n$ all the inequalities hold as equalities.} by putting $\X  \defAs  \Xj{n-1}$ and $\Z  \defAs  \Zj{n-1}$, the expression (\ref{sumsb}) gets bounded from above by
\begin{equation}
    \label{sumsc}
\sum_{j=1}^{n}\!\!\left( \expect(\Z) + \expect( \X) + \expect  \left( \sum_{h=1}^{\X}  \X \right) + \expect  \left( \sum_{h=1}^{\X}\sum_{k=1}^{\X} \Z \right) \right) .
\end{equation}

Let $Z_{i}$ and $X_h$ be the indicator variables defined for $i=1,\ldots,m$ and $h=1,\ldots,m$, respectively as
$$
\begin{array}{ll}
Z_{i} \defAs \begin{cases}
1 & \text{if } i \in \texttt{P}_{n}\\
0 & \text{otherwise}
\end{cases}
& \textrm{ and~~ }
X_{h} \defAs \begin{cases}
1 & \text{if } X \geq h\\
0 & \text{otherwise}\,,
\end{cases}
\\
\end{array}
$$
Hence
$$
\begin{array}{l}
\displaystyle \Z = \sum_{i=1}^{m}\Z_{i},\ \expect(\Z_{i}^{2}) = \expect(\Z_{i}) = \Prob\{x_{i} \trdsuff y\},\\
\displaystyle \X = \sum_{h=1}^{m}\X_{h},\  \textrm{ and }\ \expect(\X_{h}^{2}) = \expect(\X_{h}) = \Prob\{X \geq h\}.\\
\end{array}
$$
So that we have
$$
\sum_{h=1}^{\X} \X = \X\X =
    \left(\sum_{h=1}^{m} \X_{h}\right) \cdot \left(\sum_{k=1}^{m}\X_{k}\right) =   \sum_{h=1}^{m}\sum_{k=1}^{m}\X_{h}\X_{k}\,.
$$
Therefore
\begin{equation*}
\sum_{h=1}^\X \sum_{k=1}^\X \Z = \X\X\Z = \left( \sum_{h=1}^m \sum_{k=1}^{m} \X_h \X_k \right) \cdot \Z,
\end{equation*}
which yields the following upper bound for (\ref{sumsc}):
\begin{equation}
    \label{sumsd}
\sum_{j=1}^{n}\!\!\left( \expect(\Z) + \expect( \X) + \left( \expect(\Z) + 1\right) \cdot \sum_{h=1}^m \sum_{k=1}^{m}\expect\!\left(\X_h \X_k\right)  \right).
\end{equation}

To estimate each of the terms $\expect(\X_h\X_k)$ in
(\ref{sumsd}), we use the well-known Cauchy-Schwarz inequality which
in the context of expectations assumes the form
    $|\expect(UV)| \le \sqrt{\expect(U^{2})\expect(V^{2})}$\,,
for any two random variables $U$ and $V$
such that $\expect(U^{2})$, $\expect(V^{2})$ and $\expect(UV)$ are
all finite.

Then, for $1 \leq h \leq m$ and $1\leq k \leq m$, we have
\begin{equation}
    \label{expProduct}
    \expect(X_hX_k) \le \sqrt{\expect(X_h^{2})\expect(X_k^{2})}
    =\sqrt{\expect(X_h)\expect(X_k)}\,.
\end{equation}

From (\ref{expProduct}), it then follows that (\ref{sumsd}) is bounded
from above by
\begin{small}
$$
\begin{array}{l}
    \label{bound}
     \displaystyle\sum_{j=1}^{n}\left( \expect(\Z) + \expect(\X) + \left( \expect(\Z) + 1\right) \cdot \sum_{h=1}^m     \sum_{k=1}^{m}     \sqrt{\expect(X_h) \expect(X_k)} \right)\\
     \displaystyle \qquad= \sum_{j=1}^{n}\!\left( \expect(\Z)    +    \expect( \X) + \left( \expect(\Z) + 1\right) \cdot      \left(\sum_{h=1}^m\sqrt{\expect(X_h)}\right) \cdot    \left(\sum_{k=1}^{m}\sqrt{\expect(X_k)}\right) \right)\\
     \displaystyle \qquad= \sum_{j=1}^{n}\!\left( \expect(\Z)    +    \expect( \X) + \left( \expect(\Z) + 1\right) \cdot      \left(\sum_{h=1}^m\sqrt{\expect(X_h)}\right)^2  \right).
\end{array}
$$
\end{small}

To better understand (\ref{bound}), we evaluate the
expectations $\expect(\X)$ and $\expect(\Z)$ and the sum
$\sum_{h=1}^m\sqrt{\expect(X_h)}$.  To this purpose, it will be
useful to estimate also the expectations
    $\expect(X_h)=\Prob\{X\ge h\}$, for $1 \leq h \leq m$, and
    $\expect(X_k)=\Prob\{x_i \trdsuff y\}$, for $1\leq k \leq  m$.

Concerning $\expect(X_h)=\Prob\{X\ge h\}$, we reason as follows.
Since $y[n-h+1\,..\,n]$ ranges uniformly over a collection of
$\sigma^{h}$ strings and there can be at most $\min(\sigma^h, m-h+1)$
distinct factors of length $h$ in $x$, the probability $\Prob\{X\ge
h\}$ that one of them matches $y[n-h+1\,..\,n]$ is at most
$\min\left(1,\frac{m-h+1}{\sigma^h}\right)$. Hence, for  $h =
1,\ldots,m$, we have
\begin{equation}
    \label{star3}
\expect(X_h) \leq \min\left(1,\frac{m-h+1}{\sigma^h}\right).
\end{equation}

In view of (\ref{star3}),  we have:
\begin{equation}
    \label{star4}
\expect(\X)=\sum_{i=0}^m i\cdot\Prob\{\X = i\}=\sum_{i=1}^m
\Prob\{\X \ge i\} \leq \sum_{i=1}^m
\min\left(1,\frac{m-i+1}{\sigma^i}\right).
\end{equation}
Let $\overline{h}$ be the smallest integer $1\le h< m$ such that
$\frac{m-h+1}{\sigma^h} < 1$. Then, from (\ref{star4}), we have
\begin{align}
    \label{star4bis}
\expect(\X) &\leq \sum_{i=1}^{\overline{h}-1} 1 +
\sum_{i=\overline{h}}^m \frac{m-i+1}{\sigma^i} \leq \overline{h} - 1 +
(m -\overline{h} +1)\sum_{i=\overline{h}}^m \frac{1}{\sigma^i} \notag\\
& < \overline{h} - 1 + \frac{\sigma}{\sigma-1}\cdot\frac{m
-\overline{h} +1}{\sigma^{\overline{h}}} < \overline{h} - 1 +
\frac{\sigma}{\sigma-1} < \overline{h} + 1\,.
\end{align}
Since $\frac{m -(\overline{h}+1) +1}{\sigma^{\overline{h}+1}} \geq 1$,
then $\sigma^{\overline{h}+1} \leq m -(\overline{h}+1) +1 \leq m-1$,
so that
$\overline{h}+1 < \log_{\sigma} m$.
Hence from (\ref{star4bis}) and $\overline{h}+1 < \log_{\sigma} m$, we obtain
\begin{equation}
    \label{Xbound}
    \expect(\X) < \log_{\sigma} m\,.
\end{equation}
Likewise, from (\ref{star3}) and $\overline{h}+1 < \log_{\sigma} m$, we have
\begin{align}
\sum_{h=1}^m&\sqrt{\expect(X_h)}  \le
\sum_{h=1}^m\sqrt{\min\left(1, \frac{m-h+1}{\sigma^k}\right)} 
= \sum_{h=1}^{\overline{h}-1}1 + \sum_{h=\overline{h}}^m
\sqrt{\frac{m-h+1}{\sigma^h}} \notag\\
&~~~~~~\le \overline{h}-1 +
\sqrt{m-\overline{h}+1}\cdot\sum_{h=\overline{h}}^m\frac{1}{\sqrt{\sigma^h}}
< \overline{h}-1 +
\frac{\sqrt{\sigma}}{\sqrt{\sigma}-1}\cdot\sqrt{\frac{m-\overline{h}+1}{\sigma^{\overline{h}}}}
\notag\\
\label{Xbound2}
&~~~~~~< \overline{h}-1 + \frac{\sqrt{\sigma}}{\sqrt{\sigma}-1} \leq
\overline{h} + 1 < \log_{\sigma} m\,,
\end{align}
where $\overline{h}$ is defined as above.

Next we estimate $\expect(Z_i)=\Prob\{x_i \trdsuff y\}$, for $1\leq i \leq m$.
Let us denote by $\md(i)$ the number of distinct strings which have a
$utd$-match with a given string of length $i$ and whose characters are
pairwise distinct.  Then
$
\Prob\{x_{i} \trdsuff y\} \leq \md(i+1)/\sigma^{i+1}
$.
From the recursion
$$
\left\{
\begin{array}{rcl}
    \displaystyle\md(0) & = & 1\\
    \displaystyle\md(k+1) & = & \displaystyle\sum_{h=0}^{k} \md(h) + \sum_{h=1}^{\lfloor    \frac{k-1}{2}\rfloor} \md(k-2h-1) \qquad \text{(for  $k \geq 0$)}\,,
\end{array}
\right.
$$
it is not hard to see that $\md(i+1) \leq 3^{i}$, for
$i=1,2,\ldots,m$, so that we have
\begin{equation}
    \label{star6}
    \expect(\Z_{i}) = \Prob\{x_{i} \trdsuff y\} \leq    \frac{3^{i}}{\sigma^{i+1}}\,.
\end{equation}

Then, concerning $\expect(\Z)$, from (\ref{star6}) we have
\begin{equation}
    \label{Zbound}
\expect(\Z) = \expect\!\left(\sum_{i=1}^{m}\Z_{i}\right) =
\sum_{i=1}^{m}\expect(\Z_{i})
\leq \sum_{i=1}^{m}\frac{3^{i}}{\sigma^{i+1}}
< \frac{1}{\sigma} \cdot \frac{1}{1 - \frac{3}{\sigma}}
= \frac{1}{\sigma - 3} \leq 1
\end{equation}
(we recall that we have assumed $\sigma \geq 4$).

From (\ref{Zbound}), (\ref{Xbound}), and (\ref{Xbound2}), it then follows that (\ref{sumsb}) is bounded from above by $n \cdot (1+\log_{\sigma}m + 3 \log^2_{\sigma}m)$,
yielding a $\bigO(n\log^2_{\sigma}m)$ average-time complexity for our automaton based algorithm.

\section{Conclusions}\label{sec:conclusions}
In this paper we proposed a first solution for the approximate string matching problem allowing for non-overlapping translocations when factors are restricted to be adjacent. Our algorithm has a $\bigO(nm^3)$-time and $\bigO(m^2)$-space complexity in the worst case, and $\bigO(n\log^2_{\sigma})$-time complexity in the average-case.
In our future research, we plan to improve the present result, by reducing the overall worst-case time complexity of the algorithm, and to generalize our algorithm also to the case of unrestricted non-overlapping translocations.

\begin{small}
\bibliographystyle{elsarticle-num}

\end{small}

\end{document}